\documentclass[12pt,preprint]{aastex}

\shorttitle{George H. Herbig}
\shortauthors{Soderblom}

\begin{document}


\title{George Howard Herbig, 1920-2013}


\author{David R. Soderblom}

\affil{Space Telescope Science Institute, Baltimore, MD 21218\\
and\\
Center for Astrophysical Sciences, Department of Physics and Astronomy,\\
 Johns Hopkins University, Baltimore MD 21218
}









A great astronomer, George Herbig, passed away in Honolulu on October 12, 2013, at the age of 93.  His life and career were long and productive, and consistently dedicated to the careful, thorough research that earned him his reputation.  Herbig spent most of his career at Lick Observatory, first as a graduate student, and then from 1949 to 1987 as a staff member, rising through the ranks to Astronomer and becoming Professor as well when the observatory moved to the University of California at Santa Cruz in the late 1960s.  Herbig retired from UCSC in 1987 and spent the remainder of his life in Honolulu at the Institute for Astronomy of the University of Hawai`i.

Herbig was particularly known for his work on newly-formed stars in our Galaxy, but his interests broadly encompassed the life history of the Sun and Solar System and the place of the Sun among the stars.  As with many of us, this work had its beginnings in an encounter with a senior astronomer, Alfred Joy of the Mount Wilson Observatory, who was studying the variable star known as T Tauri \citep{joy45}.  Herbig was fascinated and worked on these stars associated with nebulosity in his thesis and then over his entire career.  In addition, photographs of the regions in which these stars are found led to the discovery of small, dense clouds of glowing gas.  The Mexican astronomer G. Haro was working on these objects (now known as Herbig-Haro objects) at the same time in the early 1950s, and subsequent work by them showed that these clouds often occur in matching pairs on either side of a very young star and that they were ejected by that star and can be seen to move on the sky after just a few years, a demonstration that energetic, dynamic processes were at work on human time-scales.  In a similar vein, one of these stars in nebulosity, FU Orionis, had brightened by a factor of about one hundred within just a few months and Herbig realized that type of change might be a typical short-lived phase for young stars, in that case caught through luck.  A more systematic search has uncovered more of these FUors.

Patient, comprehensive spectroscopic investigations were his forte, and in the early and mid-1950s he was able to demonstrate that the T Tauri stars and stars like them were the youngest stars in our Galaxy, no more than a few million years old.  In so doing he forged a new branch of astronomy, the study of pre-main sequence stars.  Herbig was not the only one pursuing these stars, but he dominated their observational study by piecing together the clues that could lead to no other conclusion.  For example, the very high density of stars in a cluster like NGC 2264 compared to the region around the Sun meant that stars form in concentration and then disperse, for a dense association could not plausibly arise by chance.  Prior to that it was not possible to rule out that the stars associated with nebulosity were just more ordinary stars like our Sun that happened to be passing through the clouds they were found in and being affected by them, leading to their spectroscopic peculiarities.

Herbig's path into a research career was along a steady trajectory of his own making and yet also had good fortune at critical points.  He was born on January 2, 1920, on a small island in the Ohio River, part of Wheeling, West Virginia.  His father, a tailor, died when he was six, and a few years later his mother took him to live in Los Angeles, where she had a brother.  Circumstances during the Depression were difficult for both of them, but Herbig had enough budding interest in astronomy to join the Amateur Astronomical Society of Los Angeles \citep{ghh38} and build his own telescope.   This gave him exposure to the very active astronomical scene of southern California, centered on the Mount Wilson Observatory, whose offices were in Pasadena, within sight of the telescopes atop the mountain.  

In addition to this avid interest in science, Herbig had help in going from LA's Polytechnic High School, from which he graduated in 1937, to UCLA.  Having an excellent public university nearby, let alone one with astronomers, was a resource he had the ability and ambition to take advantage of.  At the same time, Jack Preston of the amateur society helped Herbig get started at UCLA, and Frederick Leonard, Chairman of the UCLA Astronomy Department, supported him in many ways, including a recommendation that he be hired at the Griffith Observatory, an outreach-oriented facility that overlooks downtown Los Angeles.  The head guide at Griffith, George Bunton, and his wife took Herbig under their wings by providing him a job as a guide, providing resources to help Herbig stay in school during those lean years.  Herbig graduated from UCLA in 1943, and his first publication (which led to an article in the L.A. Times) appeared in 1940 \citep{ghh40}.

Herbig went to U.C. Berkeley to seek a Ph.D. in astronomy, despite having been advised at one point that he would most likely be best as a high school teacher.  While there he was assigned to the Berkeley Radiation Laboratory on a war-related project.  He found that work unsatisfying and was reassigned to the Lick Observatory, on Mount Hamilton, another branch of the University of California.  We can be confident that he found that much more to his liking, for he was to spend two decades living on Mount Hamilton and another two associated with Lick when the observatory's research staff moved to U.C. Santa Cruz.

Herbig completed his doctorate in 1948, supervised by Harold Weaver of U. C. Berkeley.  The young Herbig clearly impressed the Lick astronomers, for the Lick Director, Donald Shane, was eager to hire Herbig onto the staff as soon as his thesis was completed, urging the President of the university to make a position available.  However, Herbig was offered a one-year fellowship by the National Research Council, and took advantage of that year to go to Yerkes Observatory to work with Otto Struve, arguably the greatest of stellar spectroscopists (see Herbig 1968), and the last of the great Struve astronomical dynasty.  Struve was later to come to Berkeley and to there influence a generation of observers.  Herbig also spent a portion of that year at the McDonald Observatory, in west Texas, and in Pasadena at the Mount Wilson Observatory, working with Walter Baade.
Herbig's first year as a Lick astronomer, 1949, was to be significant in several ways.  He had married in 1943 while at UCLA, and the couple's first child was born in 1949; three more were to follow.  Most important for his future efforts, in 1949 Lick bought the 120-inch mirror blank from Palomar that would enable them to build the world's second largest optical telescope, now known as the Shane 3-meter telescope.  Despite his youth, Herbig was to take charge of the design and operation of the new telescope's coud\'{e} focus, where large optics could be installed to take spectra of high resolution.  Such facilities were rare, and so Herbig spent time visiting Mount Wilson and the coud\'{e} spectrograph of itÕs then 32-year-old 100-inch telescope, and to work with Mount Wilson astronomer Ira Bowen, who was to make the diffraction gratings for the 120-inch coud\'{e} because almost no one else had that capability.

The 120-inch opened in 1959, and its initiation and construction make an important story in itself.  That telescope helped make Lick a more significant center of astronomical research, and it also led to growth in astronomy at many of the U.C. campuses.  Great facilities inspire great science.  That growth in U.C. astronomy led to stresses between Lick and the other campuses, and Herbig led the Mount Hamilton staff in deciding to relocate to the brand new U.C. campus at Santa Cruz, a small seaside resort town that the campus was to transform.  Staying on Mount Hamilton was no longer a realistic option, while going to Santa Cruz offered the opportunity of maintaining some independence of Lick from the other U.C. campuses while establishing a new academic department, one that has continued to grow in the decades since.

Before the 120-inch was built, Lick's primary telescope was the 36-inch Crossley reflector.  After Herbig returned to Lick in 1949 he lost no time in vigorously continuing his work on the stars associated with nebulosity, and he built an instrument for the Crossley that would allow him to identify more effectively additional T Tauri stars.  A full understanding of their nature and significance required spectra of individual stars, but he also needed to know how many and where these stars were found, and that required a means to pick out the T Tauri stars over a large field of view.  T Tauri stars were identified by their having spectra roughly similar to the Sun's and cooler, as well as prominent emission in the 6563 \AA\ H$\alpha$ line of hydrogen.  Photographic emulsions of the time responded much more to blue photons than to the red ones of H$\alpha$, and so acquiring the observations meant designing and building a slitless spectrograph of high efficiency.  The emission-line stars were then reobserved spectroscopically at higher resolution.

The life of an observer was one that fit Herbig well.  In his graduate student years of the 1940s, American astronomy was still a small-scale effort at a very small number of monastic outposts and a few of the major universities.  Nearly all work was photographic, and patience and technical skill were required to get quality results.  Excellent physical insight was also necessary to interpret what was obtained, and looking back one is impressed with how well those astronomers prized the secrets of the heavens out of the observations they were able to obtain with problematic detectors (photographic plates were inefficient and non-linear) and small telescopes.

Located on Mount Hamilton, east of San Jose, California, Lick was part of the University of California, but it was not an academic department and had an arrangement with U.C. Berkeley to work with graduate students.  During that time Herbig supervised or influenced many students who went on to astronomical careers of their own, including Elizabeth Roemer, Robert Kraft, George Preston, Beverly Lynds, Ann Merchant Boesgaard, and Leonard Kuhi.  At UCSC, he supervised students under the auspices of the Board of Studies in Astronomy and Astrophysics, including Robert Zappala, William Alschuler, N. Kameswara Rao, Douglas Duncan, and Geoff Marcy, and he significantly influenced many more through the graduate courses he taught and personal interactions.  Herbig would note to his class that in physics finding the gold mostly meant great effort ``excavating'' in large groups, while in astronomy one could walk around and find the nuggets by kicking over rocks.  He never tired of exploring new veins.

I was also one of Herbig's thesis students and also a research assistant of his for several years.  We spent many nights in the 120-inch coud\'{e}, where I learned the art and craft of observing and a feel for the night sky and also what a special privilege it is to have great instruments to examine the cosmos.  There was enough time, and an interest on Herbig's part, to observe Comet Kohoutek and Comet West with our eyes on the telescope's coud\'{e} slit and to obtain good spectra.  Herbig was responsible for the coud\'{e} spectrograph, and we stopped in the middle of one dark, clear night to align optics that needed real starlight to test properly.  From all this I could see his style for being a scientist: Plan carefully, get all the data you need for the study, and make no statements not rigorously supported by the observations.

Herbig tended toward solo efforts: Of his 171 papers in refereed journals, nearly 2/3 are single author, and only about 10\% have 3 or more authors.  Yet his influence was great and well recognized.  He was awarded the Helen B. Warner Prize of the American Astronomical Society in 1955 for his work on the T Tauri stars, and in 1975 he received the AASÕ highest recognition, the Henry Russell Norris Lectureship.

Herbig's scientific career included more than just young stars.  He developed instruments and improvements to efficiency and data quality constantly, and was ready to adopt new technology as it became available to push his work further.  He did significant studies of the interstellar medium and on the enigmatic diffuse interstellar bands.  He obtained an extraordinary collection of high-quality spectra of comets, perhaps in the hope of discovering a connection between them and the interstellar medium.  And he hosted what was very likely the first ever workshop on detecting exoplanets, in Santa Cruz in 1976 \citep{green76}.

When Herbig left Santa Cruz in 1987 and went to the University of Hawai`i he took on a different role.  He no longer had direct responsibilities as he did at Lick and instead spent full time on research and working with students (including supervising Scott Dahm).  He also became someone that the younger staff and students could find a ready ear in, and he influenced many careers as he continued to study young stars in clusters and associations.

Herbig's research record will continue to speak for itself, but we will no longer have his voice of counsel.  Many of us still work by his example.







\acknowledgments

I am grateful for assistance received and conversations with Hannelore Herbig, Marilyn Wood, Bo Reipurth, Ann Merchant Boesgaard, George Preston, Joseph Wampler, Steven Vogt, Lynne Hillenbrand, and Geoff Marcy.  The Special Collections of the library at U.C. Santa Cruz was helpful in allowing access to the Mary Shane Archives of the Lick Observatory.  Roger Griffin kindly supplied a copy of the minutes from the 1976 exoplanets workshop that he and I attended.






\appendix

\section{Annotated list of selected publications}

It is not possible to present the entire panoply of Herbig's work here, so I have chosen a few particularly important research papers and some of the insights and comments he made over the years in connection with scientific symposia.  There are more, and I encourage the reader to spend a little time looking.

\subsection*{The history of the star formation field}

Herbig was intimately involved in advancing the field of star formation through his observations,  he also personally knew the key players, and he had thorough knowledge of the literature.  He has left behind several concise accounts of the state of knowledge of star formation that are enjoyable to read and provide key insights.  Because they go back some years, they must be found in bound volumes, not on-line, but are well worth the effort.  

In an introductory chapter to the Saas-Fee Advanced Course 29 (``Physics of star formation in galaxies," 1999) Herbig summarizes the history of star formation from about 1900 to the mid-1950s with the benefit of hindsight.  There are also two  accounts that were written contemporaneously.  ``On the nature and origin of the T Tauri stars'' is the transcript of a presentation at IAU Symposium 3, held in Dublin in 1955.  It shows the arguments then being made and the uncertainties that still needed to be resolved.  ``T Tauri stars, flare stars, and related objects as members of stellar associations,'' a Vatican Observatory symposium published in Ricerche Astronomische, v. 5, Specola Vaticana, 1958, tells a similar story a few years later.  In addition to again providing insight into the state of thinking at the time, it is remarkable to read through the questions and answers recorded at the end of Herbig's Vatican presentation and to realize what a extraordinary constellation of great astronomers was assembled to hear from one another, a much more diverse group than would occur today.  For additional remarks see \citet{ghh69x} in one of the Li\`{e}ge symposia.

\subsection*{Scientific insights into the advancing field}

One of Herbig's most cited publications was his 1962 review \citep{ghh62a}: ``The properties and problems of T Tauri stars and related objects.''  This review remains a must-read for anyone wanting to work in the field of star formation, and the problems described are mostly still with us.

A volume titled ``Spectroscopic astrophysics: An assessment of the contributions of Otto Struve'' was edited by Herbig and published by the University of California Press in 1970 \citep{ghh70c}.  Struve was a truly remarkable astronomer who contributed fundamentally in many ways.  Each of a number of seminal papers by Struve is coupled to a modern review of the subject that Struve's paper opened up.  My copy of this is a personal favorite.  Herbig himself contributed a chapter on T Tauri stars.

\subsection*{Research papers}

In the Struve memorial volume that he edited \citep{ghh70c}, Herbig celebrated the many pioneering studies of Struve.  Herbig himself merits such a volume because of the many seminal studies he published (and such a volume is being prepared by B. Reipurth).  As an example, Herbig's most-cited paper is his 1960 study that defined the ``Herbig AeBe stars.''  The T Tauri stars he had worked on so extensively represented mostly just low-mass stars, and the higher masses were conspicuously absent.  In \citet{ghh60b} (``The spectra of Be- and Ae-type stars associated with nebulosity'') he reported on his search for higher mass examples of pre-main sequence stars; this has now become a subfield of its own \citep[see, e.g.,][]{ghh94w}.

We now know that the Sun and stars like it rotate much more slowly than one would expect from a simple extrapolation of the trend of rotation in higher-mass stars, and that that slow rotation is the result of angular momentum loss throughout the star's life.  But \citet{ghh55d} (``Axial rotation and line broadening in stars of spectral types F0-K5'') is the first observational study that established the slow rotation of stars F8 and later, even if only upper limits were seen in most cases.  Later studies filled in once higher resolution could be achieved, but this study was the first systematic attempt.  

Similarly, \citet{ghh65e} (``Lithium abundances in F8-G5 stars'') established the basic facts regarding lithium in solar-type stars.  Abundant lithium is a defining characteristic of the T Tauri stars, and it was well known that lithium is virtually absent in integrated sunlight, yet very strong in spectra of sunspots.  Herbig showed that solar-type stars exhibit a range of lithium abundances and that that range was probably due to a steady decline in the abundance with time.  Decades later we have much more and better observations but still do not understand the behavior of lithium in stars, even though it appears that the decline in surface lithium is an indicator of convective processes.  Herbig also studied lithium in some clusters \citep{ghh65g}, and attempted to measure lithium isotope ratios \citep{ghh64h}.  All of this was enabled by high-quality spectra from the 120-inch coud\'{e} spectrograph.

Herbig studied the interstellar medium extensively, in part because the ISM feeds star formation.  He is particularly known for detailed studies of the diffuse interstellar bands.  \citet{ghh75q} is a classic instance of his very thorough approach to gathering the best observations he could and then analyzing them in detail.  The diffuse interstellar bands remain enigmatic, despite years of effort by many individuals.  When a new high-resolution spectrograph became available at the 120-inch coud\'{e} he immediately applied it to this problem \citep{ghh82q}.  \citet{ghh95x} is a comprehensive review of the diffuse interstellar bands.  \citet{ghh68t} presents an in-depth look at the interstellar lines in just one star, $\zeta$ Oph, including abundance analyses of atoms and molecules.

\citet{ghh66r} (``On the interpretation of FU Orionis'') pointed out the extraordinary behavior of FU Orionis, a T Tauri star that brightened suddenly by a factor of 100.  Such dramatic changes on short time-scales in pre-main sequence stars had not seemed possible.  That paper led to Herbig's more systematic and detailed study of the phenomenon \citep{ghh77q} (``Eruptive phenomena in early stellar evolution").

\citet{ghh77b} (``Radial velocities and spectral types of T Tauri stars'') reports on his effort over nearly a decade to assemble the quantity and quality of observational data needed to study the composition and dynamics of star-forming regions.  To make these faint stars accessible to spectra of sufficient resolution, he built a device that used an image intensifier that was optimized for the near-infrared, with photographic plates.  The image intensifier introduced distortions and Herbig developed fitting methods to extract the radial velocities in a way that kept systematics under control.  Together he and I took a thousand or more of these spectra over many, many nights.

In ``Spectral classification of faint members of the Hyades and Pleiades and the dating problem in Galactic clusters,''  \citet{ghh62w} pointed out the apparent discrepancy in ages between the main sequence turn-off at the high-mass end and the fact that low-mass stars were also on the main sequence and not displaced above it.  Some of this discrepancy has been removed by increases in the turn-off ages as the interior physics of intermediate- and high-mass stars has become better understood, but the problem remains at some level.

Finally, there are the post-T Tauri stars.  \citet{ghh73c} pointed out that the expected duration of the T Tauri phase is about 10\% of the total time needed for a star to contract to the main sequence.  By that time many T Tauri stars were known, but where were the somewhat older objects, the ``post-T Tauri stars''?  There should be many, yet few are observed.  In \citet{ghh78a} he discussed the post-T Tauri problem further, and it is a problem that is still with us.

\clearpage



\begin{figure}
\epsscale{.80}
\plotone{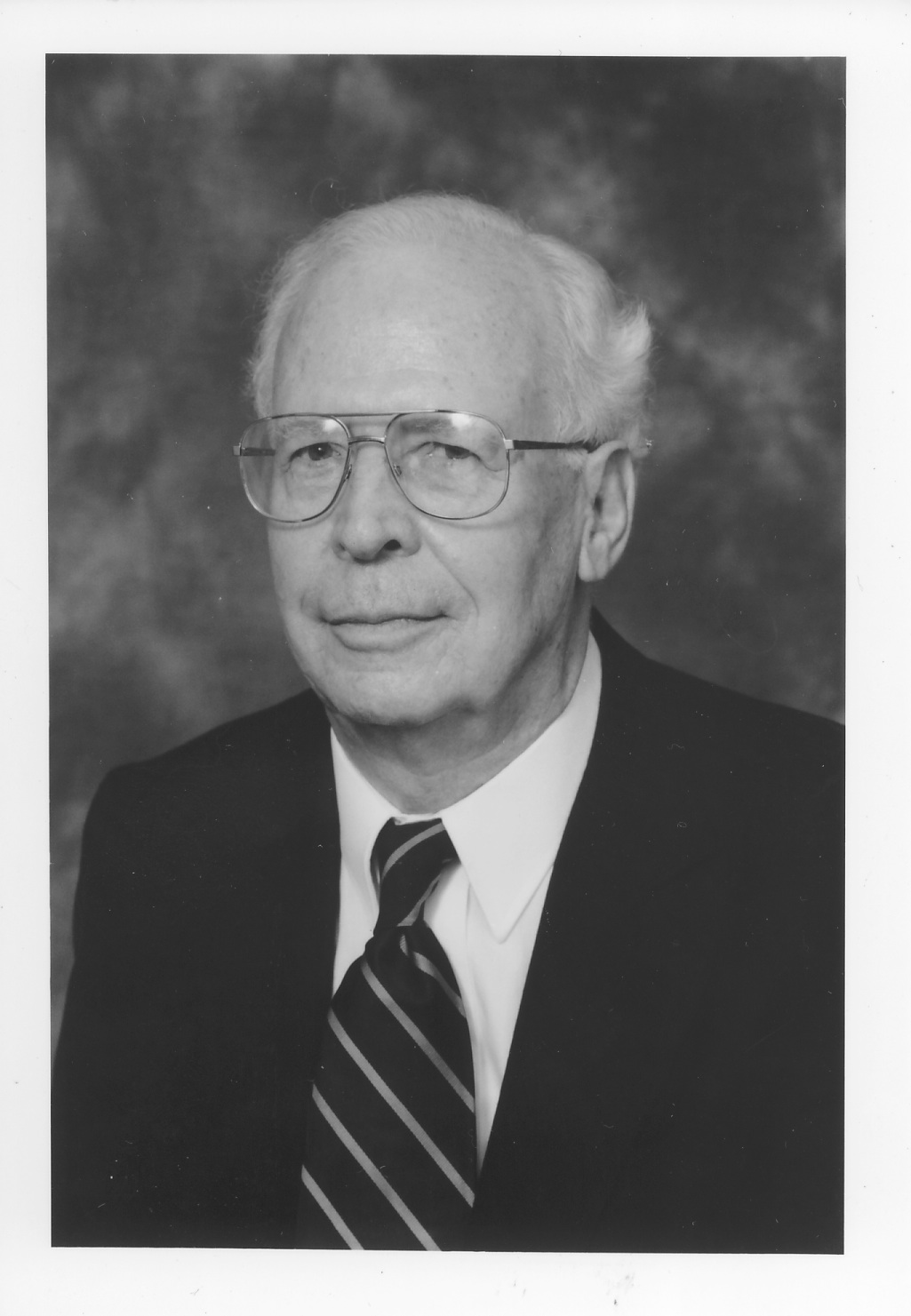}
\caption{(This should be used without a caption at the start of the obituary.)}
\end{figure}





\begin{thebibliography}{}

\bibitem[Greenstein(1976)]{green76}Greenstein, J. 1976, Minutes of ``First workshop on extrasolar planetary detection,'' personal copy of author.
\bibitem[Herbig(1938)]{ghh38}Herbig, G. H. 1938, \pasp, 46, 293
\bibitem[Herbig(1940)]{ghh40}Herbig, G. H. 1940, \pasp, 52, 327
\bibitem[Herbig(1960)]{ghh60b} Herbig, G. H. 1960, \apjs, 4, 337
\bibitem[Herbig(1962a)]{ghh62a} Herbig, G. H. 1962a, Adv. Astron. Astrophys., 1, 47
\bibitem[Herbig(1962b)]{ghh62w} Herbig, G. H. 1962b, \apj, 135, 736
\bibitem[Herbig(1964)]{ghh64h} Herbig, G. H. 1964, \apj, 140, 702
\bibitem[Herbig(1965)]{ghh65e} Herbig, G. H. 1965, \apj, 141, 588
\bibitem[Herbig(1966)]{ghh66r} Herbig, G. H. 1966, Vistas in Astron., 8, 109
\bibitem[Herbig(1968)]{ghh68t} Herbig, G. H. 1968, Z. Ap., 68, 243
\bibitem[Herbig(1969)]{ghh69x} Herbig, G. H. 1969, Mem. Soc. Roy. Sci. Li\`{e}ge, v. 19
\bibitem[Herbig(1970)]{ghh70c} Herbig, G. H. 1970, editor, {\it Spectroscopic astrophysics: An assessment of the contributions of Otto Struve}, Berkeley: University of California Press
\bibitem[Herbig(1973)]{ghh73c} Herbig, G. H. 1973, \apj, 182, 129
\bibitem[Herbig(1975)]{ghh75q} Herbig, G. H. 1975, \apj, 196, 129
\bibitem[Herbig(1977a)]{ghh77b} Herbig, G. H. 1977a, \apj, 214, 747
\bibitem[Herbig(1977b)]{ghh77q} Herbig, G. H. 1977b, \apj, 217, 693
\bibitem[Herbig(1978)]{ghh78a} Herbig, G. H. 1978, in Problems of physics and evolution of the universe, ed. L. V. Mirzoyan, Yerevan: Armenian Acad. Sci., 171
\bibitem[Herbig(1994)]{ghh94w} Herbig, G. H. 1994, in The nature and evolutionary status of Herbig Ae/Be stars, eds. P. The et al., San Francisco: ASP Press
\bibitem[Herbig(1995)]{ghh95x}Herbig, G. H. 1995, Ann. Rev. Astr. Ap., 33, 19
\bibitem[Herbig \& Soderblom(1982)]{ghh82q} Herbig, G. H., \& Soderblom, D. R. 1982, \apj, 252, 610
\bibitem[Herbig \& Spalding(1955)]{ghh55d} Herbig, G. H., \& Spalding, J. 1955, \apj, 121, 118
\bibitem[Joy(1945)]{joy45} Joy, A. H. 1945, \apj, 102, 168
\bibitem[Wallerstein et al.(1965)]{ghh65g} Wallerstein, G., Herbig, G. H., \& Conti, P. S. 1965, \apj, 141, 610


    
\end{thebibliography}
\end{document}